\documentclass[apjl]{emulateapj}

\usepackage{color}

\shorttitle{Bright points in the quiet Sun}
\shortauthors{Riethm\"uller et al.}

\begin{document}

\renewcommand{\thefootnote}{\fnsymbol{footnote}}
\renewcommand{\thempfootnote}{\fnsymbol{mpfootnote}}

\newcommand{\sunrise}{\textsc{Sunrise}}
\newcommand{\carcsec}{$\mbox{.\hspace{-0.5ex}}^{\prime\prime}$}

\title{Bright points in the quiet Sun as observed in the visible and near-UV by the balloon-borne observatory \sunrise{}}

\author{\textsc{
T.~L.~Riethm\"uller,$^{1}$
S.~K.~Solanki,$^{1,2}$
V.~Mart\'{\i}nez Pillet,$^{3}$
J.~Hirzberger,$^{1}$
A.~Feller,$^{1}$
J.~A.~Bonet,$^{3}$
N.~Bello Gonz\'alez,$^{4}$
M.~Franz,$^{4}$
M.~Sch\"ussler,$^{1}$
P.~Barthol,$^{1}$
T.~Berkefeld,$^{4}$
J.~C.~del~Toro~Iniesta,$^{5}$
V.~Domingo,$^{6}$
A.~Gandorfer,$^{1}$
M.~Kn\"olker,$^{7}$
\& W.~Schmidt$^{4}$
}}
\affil{
$^{1}$Max-Planck-Institut f\"ur Sonnensystemforschung, Max-Planck-Str. 2, 37191 Katlenburg-Lindau, Germany; riethmueller@mps.mpg.de\\
$^{2}$School of Space Research, Kyung Hee University, Yongin, Gyeonggi, 446-701, Republic of Korea\\
$^{3}$Instituto de Astrof\'{\i}sica de Canarias, C/Via L\'actea s/n, 38200 La Laguna, Tenerife, Spain\\
$^{4}$Kiepenheuer-Institut f\"ur Sonnenphysik, Sch\"oneckstr. 6, 79104 Freiburg, Germany\\
$^{5}$Instituto de Astrof\'{\i}sica de Andaluc\'{\i}a (CSIC), Apartado de Correos 3004, 18080 Granada, Spain\\
$^{6}$Grupo de Astronom\'{\i}a y Ciencias del Espacio, Universidad de Valencia, 46980 Paterna, Valencia, Spain\\
$^{7}$High Altitude Observatory, National Center for Atmospheric Research,\footnote{The National Center for Atmospheric Research is sponsored by the National Science Foundation.} P.O. Box 3000, Boulder, CO 80307-3000, USA\\
}

\begin{abstract}
   Bright points (BPs) are manifestations of small magnetic elements in the solar photosphere. Their
   brightness contrast not only gives insight into the thermal state of the photosphere (and chromosphere)
   in magnetic elements, but also plays an important role in modulating the solar total and spectral irradiance.
   Here, we report on simultaneous high-resolution imaging and spectropolarimetric observations of BPs using
   \sunrise{} balloon-borne observatory data of the quiet Sun at the disk center. BP contrasts have been
   measured between 214~nm and 525~nm, including the first measurements at wavelengths below 388~nm.
   The histograms of the BP peak brightness show a clear trend toward broader contrast distributions and higher
   mean contrasts at shorter wavelengths. At 214~nm, we observe a peak brightness of up to five times the
   mean quiet-Sun value, the highest BP contrast so far observed. All BPs are associated with
   a magnetic signal, although in a number of cases it is surprisingly weak. Most of the BPs show only weak
   downflows, the mean value being 240~m~s$^{-1}$, but some display strong down- or upflows reaching a few
   km~s$^{-1}$.
\end{abstract}


\section{Introduction}

   Bright points (BPs) are small-scale brightness enhancements located in the
   darker intergranular lanes. \citet{Dunn1973} and \citet{Mehltretter1974} were the first to describe
   these BPs in filter images taken in the far line wings of H$\alpha$ and Ca\,{\sc ii}~K,
   respectively. Common models consider BPs as radiative signatures of
   magnetic elements, which are often described by nearly vertical slender flux tubes or sheets
   \citep{Deinzer1984,Solanki1993}. The increased magnetic pressure within the flux tube leads to
   its evacuation, and the lateral inflow of heat through the walls of the flux tube makes it hot
   and bright \citep{Spruit1976}. Consequently, BPs are often used as tracers of magnetic elements.

   The contrast of BPs relative to the average quiet-Sun intensity is of interest since it provides
   insight into the structure and thermodynamics of magnetic elements. Furthermore,
   the excess brightness of magnetic elements is an important contributor to variations of the total
   solar irradiance \citep{Solanki2002}. In the visible, the contrast of BPs is particularly high
   in wavelength regions that are dominated by absorption bands of temperature-sensitive molecules
   such as CH and CN \citep{Muller1984,Berger1995,Zakharov2005,Berger2007,Utz2009}. We expect that
   their contrast is also large in the ultraviolet (UV), but they have never been studied at
   wavelengths shorter than 388~nm.

   In this work, we extend the study of BPs, in particular their contrasts, to the UV spectral range
   down to 214~nm, using seeing-free images gathered by the balloon-borne 1-m aperture \sunrise{}
   telescope. This is particularly important owing to the finding that irradiance changes below
   400~nm produce over 60\% of the variation of the total solar irradiance over the solar cycle
   \citep{Krivova2006}. The spectral and total irradiance variations are caused by the variation
   of magnetic flux at the solar surface, in particular in the form of small-scale magnetic elements
   \citep{Krivova2003,Krivova2006,Wenzler2006}.

\section{Observations, data reduction, and analysis}

   The data employed here were acquired during the 2009 June stratospheric flight of \sunrise{}. Technical
   details of the telescope are described by \citet{Barthol2010}. Image stabilization, feature tracking,
   and correction of low-order wavefront aberrations were achieved by the gondola's pointing system in
   conjunction with a six-element Shack$-$Hartmann correlating wavefront sensor \citep{Berkefeld2010}.

   Observations in the near-ultraviolet spectral domain between 214~nm and 397~nm were acquired with the
   Sunrise Filter Imager \citep[SuFI;][]{Gandorfer2010}. Simultaneously, the full Stokes magnetograph
   IMaX \citep{MartinezPillet2010} scanned the Fe\,{\sc i} line at 525.02~nm (Land\'e factor $g=3$), 
   hence providing kinematic and magnetic information. An overview of the collected data and a description
   of some of the observed phenomena is given by \citet{Solanki2010}.

   We use three time series recorded from 00:36 to 00:59~UT, 01:31 to 02:00~UT, and 14:22 to 15:00~UT
   on 2009 June 9. At these times, the telescope pointed to quiet regions close to the disk center.
   The SuFI instrument recorded filtergrams centered at 397~nm (Ca\,{\sc ii}~H), 388~nm (CN), 312~nm, 300~nm,
   and 214~nm, while IMaX was operated in its standard vector spectropolarimeter mode, i.e., full Stokes
   observations at five wavelength points with six accumulations. The five wavelength points were set to
   $\lambda-\lambda_0=-80,-40,+40,+80,$ and $+227$~m\AA~relative to the center of the line. A summary
   of the filter widths (FWHM) and exposure times can be found in Table~\ref{ExpTimes}. The effective exposure
   time of an IMaX continuum image was 6~s (6 accumulations $\times$ 4 modulation states $\times$ 250~ms), while
   Dopplergrams need five wavelength points; i.e., their effective exposure time was 30~s
   \citep[see][for more details]{MartinezPillet2010}.

   \begin{table}
   \caption{Exposure Times of the Used Time Series.}
   \label{ExpTimes}                                      
   \centering                                            
   \begin{tabular}{l l l l l}                            
   \hline                                                
   \noalign{\smallskip}
   Central     & FWHM         & 00:36$-$00:59 & 01:31$-$02:00 & 14:22$-$15:00 \\
   Wavelength  & of Filter    & Exp. Time     & Exp. Time     & Exp. Time     \\
   \noalign{\smallskip}                                                                                                                           
   (nm)        & (nm)         & (ms)          & (ms)          & (ms)          \\
   \hline                                                                                                                                            
   \noalign{\smallskip}                                                                                                                              
   \tt{214}    & \tt{10}      & \tt{...}      & \tt{...}      & \tt{30000}    \\
   \tt{300}    & \tt{~5}      & \tt{500}      & \tt{500}      & \tt{~~250}    \\
   \tt{312}    & \tt{~1.2}    & \tt{150}      & \tt{150}      & \tt{~~300}    \\
   \tt{388}    & \tt{~0.8}    & \tt{~80}      & \tt{~80}      & \tt{~~150}    \\
   \tt{396.8}  & \tt{~0.18}   & \tt{960}      & \tt{960}      & \tt{~~900}    \\
   \tt{525.02} & \tt{~0.0085} & \tt{250}      & \tt{250}      & \tt{~~250}    \\
   \noalign{\smallskip}
   \hline                                                
   \noalign{\smallskip}
   \end{tabular}
   \end{table}

   At a typical flight altitude of around 36~km, the 214~nm wavelength range was still strongly attenuated by
   the residual atmosphere of the Earth, so that an exposure time of 30~s was needed, even at the highest Sun
   elevations achieved during the flight. The 214~nm data were only acquired during the third time series.
   The cadence of the IMaX data was always 33~s, while it was 12~s for the SuFI data of the first
   two time series (i.e., all recorded SuFI wavelengths within 12~s) and it was 39~s for the last time series
   that included the 214~nm channel.

   The data were corrected for dark current and flat field. Additionally, the instrumental polarization
   was removed from the IMaX data with the help of Mueller matrices determined by pre-flight polarimetric
   calibrations. Finally, the in-flight phase-diversity measurements (permanently for SuFI and intermittently
   for IMaX) were used to correct the images for low-order wavefront aberrations \citep{Hirzberger2010a,
   MartinezPillet2010}. The phase-diversity reconstruction of the SuFI images was done using averaged
   wavefront errors \citep[level-3 data, see][]{Hirzberger2010b}. All intensity images were normalized
   to the intensity level of the mean quiet Sun, $I_{\rm{QS}}$, which was defined as the average of the whole
   image. The images were then re-sampled to the common plate scale of 0\carcsec{}0207 pixel$^{-1}$
   (original plate scale of SuFI's 300~nm images) via bilinear interpolation, and the common field of view
   (FOV) of 13\arcsec{}$\times$38\arcsec{} was determined. Residual noise was removed
   by applying a running boxcar filter of 3$\times$3 pixels. Its width corresponds to about half of the
   spatial resolution of about 0\carcsec{}15 reached by the considered observations (determined
   from radially averaged power spectra). As a proxy of the longitudinal magnetic field, the circular
   polarization degree $\langle p_{\rm{circ}} \rangle$ averaged over the four points in the line ($-80,-40,+40,+80$~m\AA),
   \[ \langle p_{\rm{circ}} \rangle = \frac{1}{4} \sum_{i=1}^{4}{a_i\frac{V_i}{I_i}}~, \]
   was calculated with a reversed sign of the two red wavelength points ($\vec{a}=[1,1,-1,-1]$) to avoid
   cancellation. The line-of-sight (LOS) component of the velocity vector was obtained by a Gaussian
   fit to the Stokes~$I$ profiles. The velocity maps were corrected for the wavelength shift over the
   FOV caused by the IMaX etalon, and a convective blueshift of 200~m~s$^{-1}$ was removed
   \citep[see][]{MartinezPillet2010}.

   Only one set of IMaX and SuFI observations every five minutes was analyzed in order to allow BPs
   to evolve between images. In total, we analyzed 19 image sets distributed over the three time series.
   BPs were manually identified in the CN images at 388~nm, in order to be consistent with earlier work that
   concentrated on visible wavelengths \citep[e.g.][]{Zakharov2005,Berger2007}. At each of the other wavelengths,
   we then determined the local brightness maximum in a 11$\times$11 pixel (i.e., 0\carcsec{}22$\times$0\carcsec{}22)
   environment of the detected BP position at 388~nm. This method takes into account that
   the various wavelengths represent different atmospheric layers (most obvious for the Ca\,{\sc ii}~H images)
   so that inclined magnetic features may appear at slightly different horizontal positions at different wavelengths.
   In total, we detected 398 BPs, of which 211 were from the third time series including the 214~nm band.

\section{Results}

\begin{figure*}
   \centering
   \includegraphics*[height=0.9\textheight]{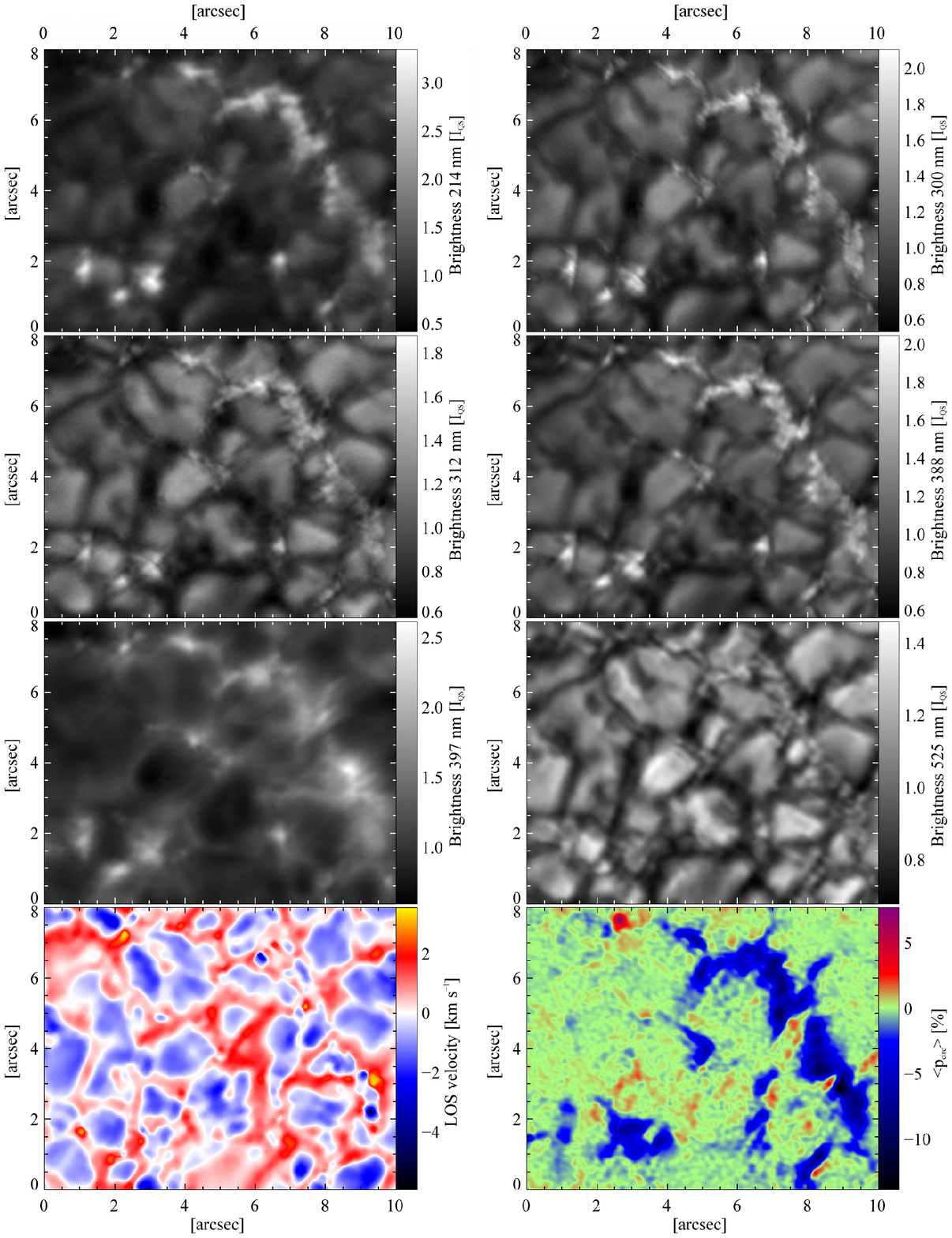}
   \caption{Intensity maps of the five wavelengths observed by SuFI (two first rows and the left panel of the third one)
   and of the continuum sample by IMaX (the right panel of the third row), all normalized to the corresponding mean
   intensity level of the quiet Sun, $I_{\rm{QS}}$. The LOS velocity obtained from a Gaussian fit (positive
   velocities correspond to downflows) and the mean circular polarization degree (see the main text for definition)
   are shown in the bottom panels. }
   \label{FigFiltergrams}
   \end{figure*}

   Figure~\ref{FigFiltergrams} shows BPs at 14:27:08 UT in a region of
   10\arcsec{}$\times$8\arcsec{} (subregion of the
   13\arcsec{}$\times$38\arcsec{} full FOV). The six upper panels display the five SuFI filtergrams and the
   IMaX Stokes~$I$ continuum image ($+227$~m\AA). Most of the bright features are visible in each
   filtergram. However, the contrast of these features at 525~nm is significantly lower than
   in the five shorter wavelengths. Some features show a prominent brightness enhancement in the Ca\,{\sc ii}~H
   image but are of moderate brightness at the other wavelengths, e.g., at position
   (8\carcsec{}6, 3\carcsec{}8). Most remarkable is the high contrast of the BPs at 214~nm,
   which strongly exceeds granulation brightness variations. It is this high contrast that lets the
   granulation appear relatively dark in the first panel of Figure 1, since the gray scales of all panels
   are adapted to the full min/max range.
   The bottom left panel shows the LOS component of the velocity. While the three BPs at the positions
   (1\carcsec{}0, 1\carcsec{}7), (1\carcsec{}8, 0\carcsec{}8), and 
   (2\carcsec{}6, 1\carcsec{}3) show strong downflows of up to 2.5~km~s$^{-1}$, most other BPs
   exhibit only moderate velocities. The bottom right panel displays the mean circular polarization degree.
   Most of the BPs in the plotted region are associated with negative polarity. Nonetheless, several bipolar
   regions can be seen, for instance the two neighboring BPs at (2\carcsec{}3, 7\carcsec{}3)
   and at (2\carcsec{}6, 7\carcsec{}7), respectively.

\begin{figure*}
   \centering
   \includegraphics*[width=\textwidth]{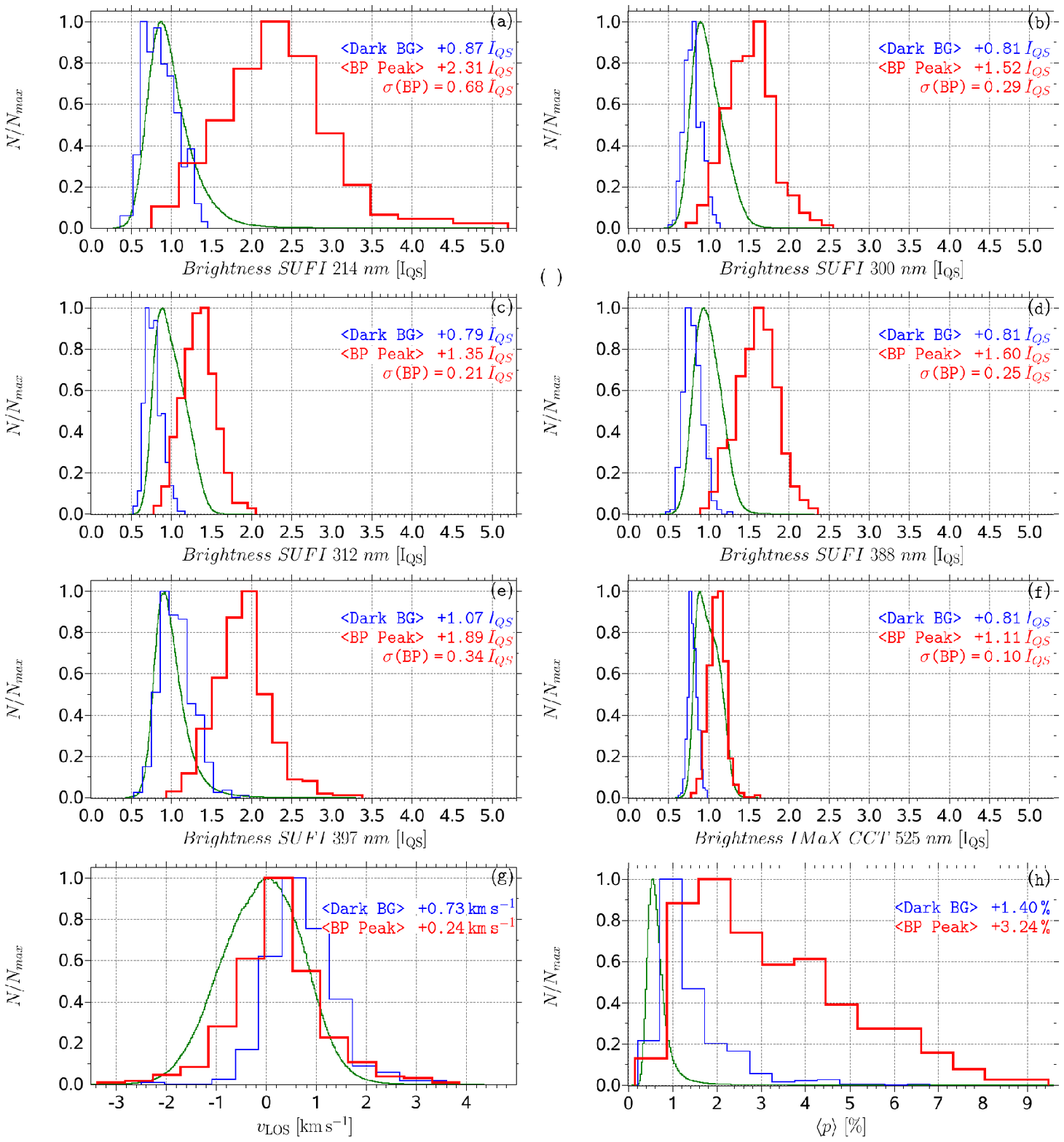}
   \caption{Brightness histograms of the six wavelengths observed by \sunrise{} together with histograms of
   the LOS velocity and the averaged polarization degree. Red lines correspond to histograms of
   the BPs' brightest pixel, blue lines to the darkest pixel in a 0\carcsec{}3 vicinity of the bright point,
   and the green lines denote histograms of all pixels in all frames. The mean values of the red and blue histograms
   as well as the standard deviations of the red brightness histograms are indicated as the text labels.}
   \label{FigHistograms}
   \end{figure*}

   Histograms of the BP peak intensity relative to the mean quiet-Sun intensity are displayed in
   Figure~\ref{FigHistograms} as red lines in panels (a)$-$(f). The blue lines correspond to the brightness
   histograms of the darkest pixel whose distance to a BP's brightest pixel is less than 0\carcsec{}3
   (typical width of an intergranular lane). The text labels denote the
   mean values. The red histograms are largely symmetric, although there is a tendency for a tail toward
   higher contrast values, in particular at 214~nm and 397~nm. As already indicated in
   Figure~\ref{FigFiltergrams}, the largest average contrast is shown by BPs in the 214~nm image. The
   highest peak brightness of 5.0~$I_{\rm{QS}}$ is also reached at this wavelength. At first sight it might
   be surprising that some BPs are less bright than some of the pixels of the blue histograms. A closer
   look reveals, however, that all BPs are indeed bright relative to the pixels in their immediate
   surroundings. For the Ca\,{\sc ii}~H line, the mean intensity of the darkest pixels
   in the BPs' vicinity is, with 1.07~$I_{\rm{QS}}$, higher than for the other wavelengths, because the Ca
   structures are generally more diffuse and larger than our limit of 0\carcsec{}3.
   The green curves give histograms of the normalized intensity
   for all pixels in all frames, practically representing the intensity
   distribution of quiet-Sun granulation. A comparison of the maximum position
   of the red and green histograms clearly shows that the BPs are much more
   conspicuous in the UV than in the visible spectral range.
   The quiet-Sun histograms are more extended toward higher intensities than
   the histograms for the BP background, since the latter largely represents
   intergranular lanes. Exceptions are the histograms for Ca\,{\sc ii}~H, which
   shows reverse granulation and those for 214~nm, where the comparable width
   indicates structures intermediate between reverse and normal granulation
   \citep[see][]{Solanki2010}.

   Panel (g) of Figure~\ref{FigHistograms} shows the LOS velocity
   histograms for the brightest BP pixels (red), the darkest pixels of their
   vicinity (blue), and for all pixels (green). On average, the LOS velocity of
   the darkest pixels shows a downflow of 730~m~s$^{-1}$, while the BPs themselves
   are associated with a clearly weaker average downflow of 240~m~s$^{-1}$. However,
   the broad wings of the distribution contain BPs with significant upflows of up
   to $-3.1$~km~s$^{-1}$ or downflows of up to 3.6~km~s$^{-1}$. 7.5\% of the
   analyzed BPs have upflows with $v_{\rm{LOS}}<-1$~km~s$^{-1}$ and 15\% of the
   BPs show downflows with $v_{\rm{LOS}}>1$~km~s$^{-1}$. The
   red histogram (BP velocities) is located between the green histogram (dominated
   by the large number of pixels showing upflows) and the blue histogram (mainly
   intergranular pixels showing downflows). Our velocities are
   obtained from a Gaussian fit to the Stokes~$I$ profiles. The Stokes~$V$
   zero-crossings may possibly show different velocities if a BP is not
   spatially resolved.

   In contrast to the last panel of Figure~\ref{FigFiltergrams}, where the signed
   net circular polarization was plotted, panel (h) of Figure~\ref{FigHistograms}
   displays the histogram of the unsigned total polarization degree $\langle p
   \rangle$ averaged over the four wavelength points in the Fe\,{\sc i} line,
   \[ \langle p \rangle = \frac{1}{4} \sum_{i=1}^{4}{\sqrt{\left( \frac{Q_i}{I_i}
   \right)^2+\left( \frac{U_i}{I_i} \right)^2+\left( \frac{V_i}{I_i} \right)^2}}~. \]
   This histogram is highly asymmetric. The polarization degree at the positions
   of the peak brightness reaches values of up to 9.1\%, the mean value is 3.24\%.
   Such large values of the polarization in 525.02~nm suggest that at least some
   of the magnetic features have been resolved \citep[see][for more details]{Lagg2010}.
   However, 4.3\% of the BPs are associated with $\langle p \rangle <1\%$.
   Although the polarization degree in the darkest surrounding pixels is weaker,
   almost all of them still show a significant polarization degree of more than
   $0.3\%$ (three times the noise level) with the strongest value being 6.6\%.
   Polarization usually displays larger structures than the BPs \citep[see the
   bottom right panel of Figure~\ref{FigFiltergrams} and ][]{Title1996},
   which can also be concluded from the fact that the blue
   histogram shows larger polarization values than the green histogram for all
   pixels.

   The mean values and their standard deviations for all red and blue histograms in Figure~\ref{FigHistograms} are
   summarized in Table~\ref{HistValues}. Columns labeled with BP refer to histograms of the BP peak value,
   while DB denotes the histograms of the dark background. The standard deviation of the BP brightness is
   $\approx0.5\left(\left\langle I/I_{\rm{QS}}\right\rangle-1\right)$ in all the SuFI channels, except Ca\,{\sc ii}~H.
   This relationship can be used to estimate the brightness histogram width for other wavelengths.
   The rightmost column of the table shows the ratio of the BP contrast (mean BP brightness value minus one)
   to the rms intensity contrast \citep[taken from][and calculated over the whole FOV of the level-3
   data]{Hirzberger2010b}. This contrast is a measure of how strongly BPs stand out relative to the
   surrounding granulation (or reversed granulation and waves in the case of 397~nm). According to
   Table~\ref{HistValues}, images at 214~nm show the BPs most clearly.

   \begin{table}
   \caption{Mean values and standard deviations of the red histograms referring to the bright points
   (labeled as BP) and blue histograms of the dark background (labeled as DB) of Figure~\ref{FigHistograms}.
   The rightmost column compares the BP contrasts with the rms granulation contrast (see main text for details).}
   \label{HistValues}                                    
   \centering                                            
   \begin{tabular}{l l l l l l}                          
   \hline                                                
   \noalign{\smallskip}
   Quantity                                    & Mean       & $\sigma$   & Mean       & $\sigma$   & Contrast  \\
                                               & BP         & BP         & DB         & DB         & Ratio     \\
   \hline                                                                                     
   \noalign{\smallskip}                                                                                                                                    
   \tt{$\left( I / I_{\rm{QS}} \right)_{214}$} & \tt{2.31}  & \tt{0.68}  & \tt{0.87}  & \tt{0.21}  & \tt{4.7}  \\
   \tt{$\left( I / I_{\rm{QS}} \right)_{300}$} & \tt{1.52}  & \tt{0.29}  & \tt{0.81}  & \tt{0.11}  & \tt{2.4}  \\
   \tt{$\left( I / I_{\rm{QS}} \right)_{312}$} & \tt{1.35}  & \tt{0.21}  & \tt{0.79}  & \tt{0.10}  & \tt{1.8}  \\
   \tt{$\left( I / I_{\rm{QS}} \right)_{388}$} & \tt{1.60}  & \tt{0.25}  & \tt{0.81}  & \tt{0.11}  & \tt{3.3}  \\
   \tt{$\left( I / I_{\rm{QS}} \right)_{397}$} & \tt{1.89}  & \tt{0.34}  & \tt{1.07}  & \tt{0.20}  & \tt{4.0}  \\
   \tt{$\left( I / I_{\rm{QS}} \right)_{525}$} & \tt{1.11}  & \tt{0.10}  & \tt{0.81}  & \tt{0.06}  & \tt{0.8}  \\
   \tt{$v_{\rm{LOS}}\,(\rm{m\,s^{-1}})$}       & \tt{240}   & \tt{910}   & \tt{730}   & \tt{650}   &           \\
   \tt{$\langle p \rangle\,(\rm{\%})$}         & \tt{3.24}  & \tt{1.82}  & \tt{1.40}  & \tt{0.84}  &           \\
   \noalign{\smallskip}
   \hline                                                
   \noalign{\smallskip}
   \end{tabular}
   \end{table}


\section{Summary and discussion}

   We identified 398 BPs in simultaneously observed photometric and polarimetric \sunrise{}
   images of a quiet-Sun region close to the disk center. Our data include three wavelengths in the near-UV
   in which the Sun was never observed before at high spatial resolution. We determined the peak brightness
   and the brightness of the dark background of every detected BP at each observed wavelength. The BPs' peak
   intensity reaches up to 5.0 times the mean quiet-Sun intensity $I_{\rm{QS}}$ at 214~nm. The mean peak intensity
   at that wavelength is 2.31~$I_{\rm{QS}}$. The 214~nm wavelength also displays the largest ratio of BP contrast
   to the rms of the intensity over the whole FOV (see the rightmost column of Table~\ref{HistValues}). This ratio
   is a measure of how prominent BPs are in an image at a particular wavelength. These values indicate that
   they are even more prominent at 214~nm than in the core of Ca\,{\sc ii}~H (for the 0.18~nm wide filter
   employed by SuFI).

   The value of the mean peak brightness at 388~nm (1.60~$I_{\rm{QS}}$) agrees exactly with the CN peak
   brightness obtained by \citet{Zakharov2005} from data recorded with the 1~m Swedish Solar Telescope
   (SST). Examples of BP brightness values (close to the disk center) given in the literature for the
   frequently observed $G$-band at 430~nm (CH molecule) are 1.45~$I_{\rm{QS}}$ \citep{Zakharov2005} and
   1.2~$I_{\rm{QS}}$ \citep{Berger2007}. Both studies analyzed SST data. \citet{Utz2009} used $G$-band data from
   the 50~cm Solar Optical Telescope onboard $Hinode$ and found 1.3~$I_{\rm{QS}}$. These values are comparable
   with what we find in the near-UV, with the exception of 214~nm, which displays distinctly higher values.
   Note that the BP contrasts given in this study are not corrected for instrumental scattered light.
   After this correction, the BP contrasts will most likely increase \citep[see, e.g.,][ A.~Feller et al. 2010,
   in preparation]{Wedemeyer2008, Mathew2009}.

   From the spectropolarimetric data we derived the LOS component of the velocity vector. The
   largest BP velocities reach $-3.1$~km~s$^{-1}$ in upflows and 3.6~km~s$^{-1}$ in downflows. The mean
   value is 240~m~s$^{-1}$, in good agreement with the average velocity of 260~m~s$^{-1}$ published
   by \citet{Beck2007} and reasonably consistent with the absence of Stokes~$V$ zero-crossing shifts
   found by \citet{Solanki1986} and \citet{MartinezPillet1997} in data with much lower spatial
   resolution. In contrast to this, \citet{GrossmannDoerth1996} reported a stronger mean downflow of
   800~m~s$^{-1}$ as derived from the zero-crossing of their Stokes~$V$ profiles and \citet{Sigwarth1999}
   found a velocity range of $\pm$5~km~s$^{-1}$ and a mean velocity of 500~m~s$^{-1}$. The nature of the
   BPs displaying strong up- or downflows in Stokes~$I$ will be investigated in a subsequent study.

   The polarization degree, averaged over the four points within the Fe\,{\sc i} line, is also calculated
   from the IMaX data and shows values up to 9.1\%. The mean BP polarization degree is 3.24\%, which
   is clearly above the mean signal of 1.40\% for the dark vicinity. Intriguingly, about 4\% of the
   BPs are associated with relatively weak magnetic flux ($\langle p \rangle <1\%$). This raises interesting
   questions, since enhanced temperatures and hence brightness in magnetic elements (flux tubes) is
   caused by evacuation, which in turn is proportional to $B^{2}$. Therefore, we would expect that BPs show
   strong Stokes~$V$ signals. Significantly inclined magnetic fields cannot explain the observed weak
   Stokes~$V$ signals, because we also measured weak Stokes~$Q$ and $U$ signals. Additionally, strongly
   evacuated flux tubes are strongly buoyant and hence should be nearly vertical \citep{Schuessler1986}.
   \citet{Lagg2010} find a strong Stokes~$I$ line weakening for kilo-Gauss network patches due to
   temperature enhancements in the flux tubes. Such an absorption weakening can also lead to a weak
   Stokes~$V$ signal. Although the network patches analyzed by \citet{Lagg2010} are spatially resolved,
   we find many BPs that exhibit complex Stokes profiles, which is a clear indication that not all of
   the BPs are spatially resolved. Insufficient resolution can also contribute to weak Stokes~$V$ signals.

   In summary, we find high intensity contrasts of BPs in the near-UV range (including the first
   measurements below 388~nm), with values up to 5~$I_{\rm{QS}}$ at 214~nm. The simultaneous spectropolarimetric
   measurements confirm the close association of BPs with magnetic flux concentrations in intergranular
   downflow lanes. However, the majority of the BPs exhibit only weak downflows.

   The reasonably high cadence of \sunrise{} data (between 4~s and 39~s, depending on the number
   of observed wavelengths and their exposure times) and the high measured contrasts of
   BPs make detailed future studies of the dynamical properties of BPs very promising
   (e.g., S.~Jafarzadeh et al. 2010, in preparation). Of considerable additional benefit for such studies will
   be the possibility to compare the dynamics at different layers in the solar atmosphere, which are covered
   by the combination of SuFI and IMaX wavelength bands.

   \begin{acknowledgements}
   The German contribution to \sunrise{} is funded by the Bundesministerium
   f\"{u}r Wirtschaft und Technologie through Deutsches Zentrum f\"{u}r Luft-
   und Raumfahrt e.V. (DLR), Grant No. 50~OU~0401, and by the Innovationsfond of
   the President of the Max Planck Society (MPG). The Spanish contribution has
   been funded by the Spanish MICINN under projects ESP2006-13030-C06 and
   AYA2009-14105-C06 (including European FEDER funds). The HAO contribution was
   partly funded through NASA grant number NNX08AH38G. This work has been partly
   supported by the WCU grant (No. R31-10016) funded by the Korean Ministry of
   Education, Science \& Technology.
   \end{acknowledgements}



\begin{thebibliography}{}

   \bibitem[Barthol et al.(2010)]{Barthol2010} Barthol, P., et al. 2010,
      Sol. Phys., in press (arXiv:1009.2689)

   \bibitem[Beck et al.(2007)]{Beck2007} Beck, C., Bellot Rubio, L.~R., Schlichenmaier, R., \& S\"utterlin, P. 2007,
      A\&A, 472, 607

   \bibitem[Berger et al.(2007)]{Berger2007} Berger, T.~E., Rouppe van der Voort, L., \& L\"ofdahl, M. 2007,
      ApJ, 661, 1272

   \bibitem[Berger et al.(1995)]{Berger1995} Berger, T.~E., Schrijver, C.~J., Shine, R.~A., Tarbell, T.~D., Title, A.~M., \& Sharmer, G. 1995,
      ApJ, 454, 531

   \bibitem[Berkefeld et al.(2010)]{Berkefeld2010} Berkefeld, T., et al. 2010,
      Sol. Phys., in press (arXiv:1009.3196)

   \bibitem[Deinzer et al.(1984)]{Deinzer1984} Deinzer, W., Hensler, G., Sch\"ussler, M., \& Weisshaar, E. 1984,
      A\&A, 139, 426

   \bibitem[Dunn \& Zirker(1973)]{Dunn1973} Dunn, R.~B., \& Zirker, J.~B. 1973,
      Sol. Phys., 33, 281  

   \bibitem[Gandorfer et al.(2010)]{Gandorfer2010} Gandorfer, A., et al. 2010,
      Sol. Phys., in press (arXiv:1009.1037)

   \bibitem[Grossmann-Doerth et al.(1996)]{GrossmannDoerth1996} Grossmann-Doerth, U., Keller, C.~U., \& Sch\"ussler, M. 1996,
      A\&A, 315, 610

   \bibitem[Hirzberger et al.(2010a)]{Hirzberger2010a} Hirzberger, J., et al. 2010a,
      A\&A, submitted

   \bibitem[Hirzberger et al.(2010b)]{Hirzberger2010b} Hirzberger, J., et al. 2010b,
      ApJ, this issue

   \bibitem[Krivova et al.(2003)]{Krivova2003} Krivova, N.~A., Solanki, S.~K., Fligge, M., \& Unruh, Y.~C.\ 2003,
      A\&A, 399, L1

   \bibitem[Krivova et al.(2006)]{Krivova2006} Krivova, N.~A., Solanki, S.~K., \& Floyd, L.\ 2006,
      A\&A, 452, 631

   \bibitem[Lagg et al.(2010)]{Lagg2010} Lagg, A., et al. 2010,
      ApJ, this issue

   \bibitem[Mart\'{\i}nez Pillet et al.(1997)]{MartinezPillet1997} Mart\'{\i}nez Pillet, V., Lites, B.~W., \& Skumanich, A. 1997,
      ApJ, 474, 810

   \bibitem[Mart\'{\i}nez Pillet et al.(2010)]{MartinezPillet2010} Mart\'{\i}nez Pillet, V., et al. 2010,
      Sol. Phys., in press (arXiv:1009.1095)

   \bibitem[Mathew et al.(2009)]{Mathew2009} Mathew, S.~K., Zakharov, V., \& Solanki, S.~K. 2009,
      A\&A, 501, L19

   \bibitem[Mehltretter(1974)]{Mehltretter1974} Mehltretter, J. P. 1974,
      Sol. Phys., 38, 43  

   \bibitem[Muller \& Roudier(1984)]{Muller1984} Muller, R., \& Roudier, Th. 1984,
      Sol. Phys., 94, 33  

   \bibitem[Sch{\"u}ssler(1986)]{Schuessler1986} Sch{\"u}ssler, M.\ 1986,
      Small Scale Magnetic Flux Concentrations in the Solar Photosphere,
      ed. W. Deinzer, M. Kn\"olker, \& H.~H. Voigt
      (G\"ottingen: Vandenhoeck \& Ruprecht), 103

   \bibitem[Sigwarth et al.(1999)]{Sigwarth1999} Sigwarth, M., Balasubramaniam, K.~S., Kn\"olker, M., \& Schmidt, W. 1999,
      A\&A, 349, 941

   \bibitem[Solanki(1986)]{Solanki1986} Solanki, S.~K. 1986,
      A\&A, 168, 311

   \bibitem[Solanki(1993)]{Solanki1993} Solanki, S.~K. 1993,
      Space Sci. Rev., 63, 1

   \bibitem[Solanki \& Fligge(2002)]{Solanki2002} Solanki, S.~K., \& Fligge, M. 2002,
      Adv. Space Res., 29, 1933

   \bibitem[Solanki et al.(2010)]{Solanki2010} Solanki, S.~K., et al. 2010,
      ApJ, this issue

   \bibitem[Spruit (1976)]{Spruit1976} Spruit, H.~C. 1976,
      Sol. Phys., 50, 269  

   \bibitem[Title \& Berger(1996)]{Title1996} Title, A.~M., \& Berger, T.~E. 1996,
      ApJ, 597, L173

   \bibitem[Utz et al.(2009)]{Utz2009} Utz, D., Hanslmeier, A., M\"ostl, C., Muller, R., Veronig, A., \& Muthsam, H. 2009,
      A\&A, 498, 289

   \bibitem[Wedemeyer-B\"ohm (2008)]{Wedemeyer2008} Wedemeyer-B\"ohm, S. 2008,
      A\&A, 487, 399

   \bibitem[Wenzler et al.(2006)]{Wenzler2006} Wenzler, T., Solanki, S.~K., Krivova, N.~A., \& Fr{\"o}hlich, C.\ 2006,
      \aap, 460, 583 

   \bibitem[Zakharov et al.(2005)]{Zakharov2005} Zakharov, V., Gandorfer, A., Solanki, S.~K., \& L\"ofdahl, M. 2005,
      A\&A, 437, L43

\end{thebibliography}
\end{document}